\documentclass[aps,pra,reprint,letterpaper,groupedaddress,floatfix,superscriptaddress]{revtex4-2}

\usepackage{bm} 
\usepackage{amsmath}
\usepackage{amssymb}
\usepackage{mathtools}
\usepackage{braket}
\usepackage{ascmac}
\usepackage{siunitx}
\usepackage{natbib}
\usepackage{multirow}

\usepackage{hyperref}
\usepackage{color}
\usepackage{graphicx}

\usepackage{here}
\usepackage{comment}
\usepackage{dcolumn}
\usepackage{threeparttable}
\DeclareSIUnit{\cps}{cps}

\makeatletter
\let\MYcaption\@makecaption
\makeatother
\usepackage[singlelinecheck=false, justification=raggedright, labelformat=simple, hypcap=true]{subcaption}

\makeatletter
\let\@makecaption\MYcaption
\makeatother

\graphicspath{{./figures}}

\newcommand{\red}[1]{\textcolor{black}{#1}}

\begin{document}


\title{Boosting the generation rate of squeezed single-photon states\\by generalized photon subtraction}


\author{Hiroko Tomoda}

\affiliation{Department of Applied Physics, School of Engineering, The University of Tokyo, 7-3-1 Hongo, Bunkyo-ku, Tokyo 113-8656, Japan}

\author{Akihiro Machinaga}
\affiliation{Department of Applied Physics, School of Engineering, The University of Tokyo, 7-3-1 Hongo, Bunkyo-ku, Tokyo 113-8656, Japan}

\author{Kan Takase}
\affiliation{Department of Applied Physics, School of Engineering, The University of Tokyo, 7-3-1 Hongo, Bunkyo-ku, Tokyo 113-8656, Japan}
\affiliation{Optical Quantum Computing Research Team, RIKEN Center for Quantum Computing, 2-1 Hirosawa, Wako, Saitama 351-0198, Japan}

\author{Jun Harada}
\affiliation{Department of Applied Physics, School of Engineering, The University of Tokyo, 7-3-1 Hongo, Bunkyo-ku, Tokyo 113-8656, Japan}

\author{Takahiro Kashiwazaki}
\affiliation{NTT Device Technology Labs, NTT Corporation, 3-1, Morinosato Wakamiya, Atsugi, Kanagawa 243-0198, Japan}

\author{Takeshi Umeki}
\affiliation{NTT Device Technology Labs, NTT Corporation, 3-1, Morinosato Wakamiya, Atsugi, Kanagawa 243-0198, Japan}

\author{Shigehito Miki}
\affiliation{Advanced ICT Research Institute, National Institute of Information and Communications Technology, 588-2 Iwaoka, Nishi-ku, Kobe, Hyogo 651-2492, Japan}

\author{Fumihiro China}
\affiliation{Advanced ICT Research Institute, National Institute of Information and Communications Technology, 588-2 Iwaoka, Nishi-ku, Kobe, Hyogo 651-2492, Japan}

\author{Masahiro Yabuno}
\affiliation{Advanced ICT Research Institute, National Institute of Information and Communications Technology, 588-2 Iwaoka, Nishi-ku, Kobe, Hyogo 651-2492, Japan}

\author{Hirotaka Terai}
\affiliation{Advanced ICT Research Institute, National Institute of Information and Communications Technology, 588-2 Iwaoka, Nishi-ku, Kobe, Hyogo 651-2492, Japan}

\author{Daichi Okuno}
\affiliation{Department of Applied Physics, School of Engineering, The University of Tokyo, 7-3-1 Hongo, Bunkyo-ku, Tokyo 113-8656, Japan}

\author{Shuntaro Takeda}
\email[]{takeda@ap.t.u-tokyo.ac.jp}
\affiliation{Department of Applied Physics, School of Engineering, The University of Tokyo, 7-3-1 Hongo, Bunkyo-ku, Tokyo 113-8656, Japan}


\date{\today}

\begin{abstract}
  In optical quantum information processing with continuous variables, optical non-Gaussian quantum states are essential for universal and fault-tolerant quantum computation.
  Experimentally, their most typical generation method is photon subtraction (PS), where single-photon detection by an on/off detector probabilistically heralds the generation of squeezed single-photon states.
  In PS, however, trying to avoid unwanted multiphoton detection inevitably limits the generation rate, hindering the application of squeezed single-photon states.
  Here, we theoretically show that generalized photon subtraction (GPS), a simple extension of PS, can improve the generation rate while maintaining the quality of the generated states.
  Furthermore, we experimentally demonstrate the generation rate improvement for 2-dB- and 4-dB-squeezed single-photon states compared to PS, by more than one order of magnitude, particularly for the case of \SI{2}{\deci\bel}.
  Our results will accelerate the application of squeezed single-photon states to more advanced quantum information protocols.
\end{abstract}


\maketitle



\section{Introduction\label{sec:Introduction}}
Thus far, quantum information processing (QIP) has been implemented with various physical systems, but optical systems are one of the reasonable choices due to their high-speed operation at room temperature~\cite{flamini_photonic_2019}.
In particular, the approach based on optical continuous variables (CVs)~\cite{andersen_continuous-variable_2010,takeda_toward_2019,asavanant_optical_nodate} offers advantageous technologies for QIP, such as deterministic entanglement generation~\cite{menicucci_universal_2006,asavanant_generation_2019} and quantum gates~\cite{yoshikawa_demonstration_2007,miwa_exploring_2014,enomoto_programmable_2021,sakaguchi_nonlinear_2023}.
Moreover, optical CV QIP can potentially implement fault-tolerant QIP in a hardware-efficient way by bosonic codes~\cite{cochrane_macroscopically_1999,gottesman_encoding_2001,jeong_efficient_2002,ralph_quantum_2003,lund_fault-tolerant_2008}, which encode one logical qubit into only one mode by utilizing the infinite-dimensional Hilbert space of harmonic oscillators.

In such optical CV QIP, non-Gaussian states are essential as auxiliary states for universal quantum computation~\cite{lloyd_quantum_1999}, quantum error correction~\cite{niset_no-go_2009}, entanglement distillation~\cite{eisert_distilling_2002,fiurasek_gaussian_2002,giedke_characterization_2002}, and other applications.
Moreover, they are used as bosonic qubits for fault-tolerant CV QIP~\cite{cochrane_macroscopically_1999,gottesman_encoding_2001,jeong_efficient_2002,ralph_quantum_2003,lund_fault-tolerant_2008}.
Optical non-Gaussian states have been experimentally generated by the heralding schemes with a photon detector, and the typical generation method is the probabilistic generation of squeezed single-photon states with photon subtraction (PS) \red{based on a beam splitter}~\cite{dakna_generating_1997}.
\red{
Although photon subtraction can be performed deterministically based on atomic systems~\cite{pinotsi_single_2008,honer_artificial_2011,rosenblum_extraction_2016,stiesdal2021controlled}, the probabilistic scheme has been more widely used for its simplicity in implementation.
}

In this scheme, we tap squeezed states by a beam splitter (BS), and detecting a reflected single photon heralds the squeezed single-photon states in the transmission side (Fig.~\ref{fig:SchematicPS}).
These squeezed single-photon states are approximate states of small-amplitude Schr\"{o}dinger's cat states and have applications in quantum computation and quantum metrology~\cite{gilchrist_schrodinger_2004}.
Moreover, such PS underlies the generation protocols of large-amplitude Schr\"{o}dinger's cat states~\cite{lund_conditional_2004,laghaout_amplification_2013,sychev_enlargement_2017} and Gottesman-Kitaev-Preskill (GKP) states~\cite{vasconcelos_all-optical_2010,weigand_generating_2018}, thus playing a key role in fault-tolerant CV QIP~\cite{cochrane_macroscopically_1999,gottesman_encoding_2001,jeong_efficient_2002,ralph_quantum_2003,lund_fault-tolerant_2008}.

However, in the typical PS with an on/off detector~\red{\cite{wenger_non-gaussian_2004,ourjoumtsev_generating_2006,neergaard-nielsen_generation_2006,wakui_photon_2007,dong_generation_2014,asavanant_generation_2017,takase_2022_generation}}, the tap ratio should be set nearly zero to reduce the probability of undesired multiphoton detection, which cannot be distinguished from desired single-photon detection.
As a result, the photon detection rate, namely the generation rate of the squeezed single-photon states, is limited.
Recently, GKP states were generated from two squeezed single-photon states~\cite{konno_logical_2024} based on the generation protocol in Refs.~\cite{vasconcelos_all-optical_2010,weigand_generating_2018}, but the generation rate was limited to only \SI{10}{\hertz} at a single-step breeding operation, whose extension to multiple steps is now required.
In order to overcome such limitations, improving the generation rate of squeezed single-photon states is essential.
For this purpose, several generation methods have been theoretically proposed~\cite{zhang_photon_2024,marek_loop-based_2018} and one method to use photon addition, instead of PS, was experimentally demonstrated recently~\cite{chen_generation_2023}.
However, the demonstrated method requires in-line optical nonlinear crystal acting on squeezed states, inevitably introducing a coupling loss on the fragile squeezed states.

In this paper, we theoretically and experimentally demonstrate the generation rate improvement by adopting generalized photon subtraction (GPS)~\cite{takase_generation_2021}.
GPS is extended from PS by adding one orthogonally squeezed input state to the PS configuration (Fig.~\ref{fig:SchematicGPS}).
In GPS with the on/off detector, we find that this additional squeezing parameter enables us to more efficiently generate the squeezed single-photon states with comparable quality to PS.
To analytically reveal that, we first model PS and GPS and derive the qualities and generation rates of the heralded states in both cases.
We then quantitatively evaluate how much GPS can increase the generation rate while maintaining the same quality.
In addition, we experimentally confirm the validity of such calculations for the generation of 2-dB- and 4-dB-squeezed single-photon states.
Their generation rates are significantly boosted, by more than one order of magnitude particularly for \SI{2}{\deci\bel}.
These results will accelerate the applications of squeezed single-photon states, which have been limited in PS. 

This paper is organized as follows.
In Sec.~\ref{sec:TheoryNumericalresults}, we theoretically analyze both PS and GPS in each case of using a photon-number resolving detector (PNRD) and an on/off detector.
As for the case with the on/off detector, we mathematically clarify the relationship between the quality and generation rate, and numerically evaluate the generation rate improvement by GPS compared with PS.
In Sec.~\ref{sec:Experimentalsetup}, we describe our experimental setup applicable to both PS and GPS.
In Sec.~\ref{sec:ExperimentalResultsDiscussion}, we first generate various quantum states to present the general characteristics of GPS and verify our theoretical model.
Next, we generate the squeezed single-photon states with the same squeezing level by both PS and GPS to experimentally validate our theoretical conclusion about the generation rate improvement.
In Sec.~\ref{sec:Conclusion}, we summarize our discussion.

\section{Theory \& Numerical results\label{sec:TheoryNumericalresults}}

\begin{figure}[tbp]
  \begin{subfigure}[t]{\linewidth}
    \centering
    \subcaption{
      \label{fig:SchematicPS}
      }
      \includegraphics[width=\linewidth]{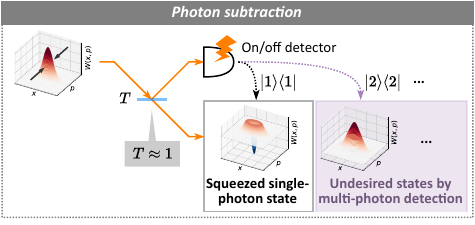}
  \end{subfigure}
  \begin{subfigure}[t]{\linewidth}
    \centering
    \subcaption{
      \label{fig:SchematicGPS}
      }
    \includegraphics[width=\linewidth]{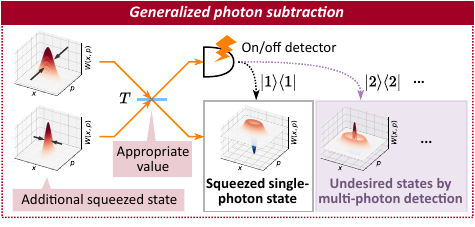}
  \end{subfigure}
  \caption{
    Generation method of squeezed single-photon states.
    (a) Photon subtraction (PS). (b) Generalized photon subtraction (GPS).
    Both PS and GPS in these two subfigures use an on/off detector causing the trade-off, which we discuss in this paper.
    \label{fig:Schematic}
  }
\end{figure}

Figure~\ref{fig:Schematic} is the configuration of PS and GPS discussed in this paper.
\red{Both PS and GPS are based on the same idea: squeezed or vacuum states (even photon-number states) are mixed at the beginning, and then single-photon detection on one side heralds squeezed single-photon states (odd photon-number states) on the other side because of the photon-number conservation.}
It has been known that PS with an on/off detector has a trade-off relationship between the quality and generation rate of the heralded states~\cite{suzuki_analysis_2006}.
In this section, we theoretically show the trade-off improvement by adopting GPS.

First, we qualitatively explain how the trade-off arises.
In PS of Fig.~\ref{fig:SchematicPS}, since an on/off detector cannot resolve the number of photons, it heralds the mixed states of single- and multiphoton subtracted states.
Hence, the transmissivity $T$ of the BS should be set to $T\approx1$ to reduce contamination of the quantum states by unwanted multiphoton detection.
However, this condition inevitably decreases the photon detection rate, corresponding to the generation rate.
On the other hand, we can increase the photon detection rate by decreasing $T$, while the quality of the squeezed single-photon state gets worse due to the larger multiphoton detection effect in this case. 
Thus there is a trade-off relationship between the quality and generation rate in PS.
In contrast, in GPS of Fig.~\ref{fig:SchematicGPS}, the additional orthogonally squeezed state realizes generating the squeezed single-photon states with the comparable quality to PS even by using lower transmissivity.
As a result, more photons reach the on/off detector, and thereby the trade-off is improved.

In the following, we theoretically analyze the qualities and generation rates in both PS and GPS and finally discuss the trade-off difference between these two generation methods.

\subsection{PS and GPS with PNRD\label{sec:PSandGPSwithPNRD}}
Before considering PS and GPS with an on/off detector, we model the simpler case with a PNRD to derive the output states for various experimental parameters.
This section is based on the more general discussion in Ref.~\cite{takase_generation_2021}.

\begin{figure}[tbp]
  \centering
  \includegraphics[width=0.9\linewidth]{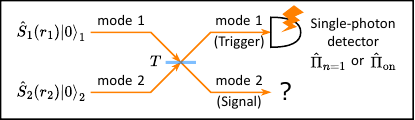}
  \caption{Schematic picture of our model.
  The projection operator of the single-photon detector depends on its detector type: the PNRD or the on/off detector.
  \label{fig:TheoreticalFigure}}
\end{figure}

Let us start the discussion using the model as illustrated in Fig.~\ref{fig:TheoreticalFigure}.
We use $\hbar=1$ throughout this paper, and define quadrature operators of two modes in Fig.~\ref{fig:TheoreticalFigure} as $\hat{x}_1$, $\hat{x}_2$, $\hat{p}_1$, and $\hat{p}_2$.
A two-mode covariance matrix $V$ is introduced by using $\hat{\bm{q}} = \left( \hat{x}_1 , \hat{x}_2, \hat{p}_1, \hat{p}_2 \right) ^{\top}$:
\begin{gather}
  V_{ij} = \frac{1}{2}\braket{\hat{q}_{i}\hat{q}_{j} + \hat{q}_{j}\hat{q}_{i}} - \braket{\hat{q}_{i}}\braket{\hat{q}_{j}}
  .
\end{gather}
We also define a squeezing operator for mode~$i$ ($i=1,2$) as $\hat{S}_i ^{\dag}(r)\hat{x}_i \hat{S}_i(r)=\hat{x}_ie^{-r}$, $\hat{S}_i ^{\dag}(r)\hat{p}_i \hat{S}_i(r)=\hat{p}_i e^{r}$ and write an $n$-photon state in mode~$i$ as $\ket{n}_{i}$.

First, we prepare two squeezed vacuum states $\hat{S}_1(r_1)\ket{0}_1$ and $\hat{S}_2(r_2)\ket{0}_2$, which are characterized by a two-mode covariance matrix $V_{0}$ as
\begin{align}
  V_{0} = 
  \text{diag}\left(\frac{e^{-2r_1}}{2}, \frac{e^{-2r_2}}{2}, \frac{e^{2r_1}}{2}, \frac{e^{2r_2}}{2}\right)
  .
\end{align}
In general, any value for two squeezing parameters $r_1$ and $r_2$ can be chosen in GPS, and PS corresponds to GPS where $r_1$ or $r_2$ is equal to zero.
However, for later convenience, we here define GPS as the case with $r_1 r_2 <0$ and PS as the case with $r_2=0$.

This state is then input to a BS with a transmissivity of $T=1-R$, whose $4\times4$ transformation matrix $U$ is written as
\begin{align}
  U & = 
  \begin{pmatrix}
    B & 0\\
    0 & B\\
  \end{pmatrix}
  ,\quad
  B = 
  \begin{pmatrix}
    \sqrt{R} & \sqrt{T}\\
    -\sqrt{T} & \sqrt{R}\\
  \end{pmatrix}
  .
\end{align}
By using the above, the covariance matrix $V_{\text{G}}$ of the BS output, which is a two-mode Gaussian state, is calculated as 
\begin{gather}
  V_{\text{G}} = U V_{0} U^{\top}
  =
  \begin{pmatrix}
    \sigma_x & 0 \\
    0 & \sigma_p\\
  \end{pmatrix}.
  \label{eq:VG}
\end{gather}
Here $2\times2$ matrices $\sigma_x$ and $\sigma_p$ are
\begin{align}
  \sigma_x &= \frac{1}{2}
  \begin{pmatrix}
    Re^{-2r_1} + Te^{-2r_2} & \sqrt{RT}(-e^{-2r_1}+e^{-2r_2}) \\
    \sqrt{RT}(-e^{-2r_1}+e^{-2r_2}) & Te^{-2r_1} + Re^{-2r_2} \\
  \end{pmatrix},
  \label{eq:definitionsigma_x}
  \\
  \sigma_p &= \frac{1}{2}
  \begin{pmatrix}
    Re^{2r_1} + Te^{2r_2} & \sqrt{RT}(-e^{2r_1}+e^{2r_2}) \\
    \sqrt{RT}(-e^{2r_1}+e^{2r_2}) & Te^{2r_1} + Re^{2r_2} \\
  \end{pmatrix}, \hspace{-5mm}
  \label{eq:definitionsigma_p}
\end{align}
which satisfy the relationship: $\sigma_x \sigma_p = I/4$ ($I$ is an identity matrix).
Moreover, this BS output can be represented with a two-mode Wigner function $W_{\text{G}}(\bm{q})$ where $\bm{q}=(x_1, x_2, p_1, p_2)^{\top}$ are quadratures.
Given zero mean value for each quadrature throughout the process, $W_{\text{G}}(\bm{q})$ is written~\cite{weedbrook_gaussian_2012} as
\begin{gather}
  W_{\text{G}}(\bm{q}) =
  \frac{1}{\pi^2}
  \exp{\left[
    -\frac{1}{2} \bm{q}^{\top} V_{\text{G}}^{-1} \bm{q}
  \right]}.
  \label{eq:Wout}
\end{gather}
Here one of the BS output (mode~1) goes to the photon detector, which we call \textit{trigger mode}.
Then, photon detection heralds the quantum state in the other side (mode~2), which we call \textit{signal mode}.

Let us discuss the state heralded by the PNRD.
The projection operator of single-photon detection on mode~1 $\hat{\Pi}_{n=1}$ is
\begin{align}
  \hat{\Pi}_{n=1} = \ket{1} \!{}_1 {}_1\!\bra{1}.
\end{align}
When the single-photon detection occurs in mode~1, the density matrix of the heralded state in mode~2 $\hat{\rho}_{\text{out}}$ is written using that of the BS output $\hat{\rho}_{\text{G}}$ as
\begin{align}
  \hat{\rho}_{\text{out}} &= 
  \frac{\text{Tr}_1 \left( \hat{\Pi}_{n=1} \hat{\rho}_{\text{G}} \hat{\Pi}_{n=1}\right)}
  {P_{n=1}},
\end{align}
where $\text{Tr}_1$ represents partial trace over mode~1 and single-photon detection probability $P_{n=1}$ is
\begin{align}
  P_{n=1} &= \text{Tr} \left( \hat{\Pi}_{n=1} \hat{\rho}_{\text{G}} \right). \label{eq:P1}
\end{align}
Then, using the following Wigner function of $\ket{n}$:
\begin{gather}
  \begin{gathered}
    W_n(x, p) = \frac{(-1) ^n}{\pi} e^{-x^2-p^2} L_n\left(2(x^2 + p^2)\right), \\
    L_n(x) = \sum_{k=0}^{n}
    \begin{pmatrix}
      n\\
      k\\  
    \end{pmatrix}
    \frac{(-1)^k}{k!}x^k\\
    (L_0(x) = 1, \ L_1(x) = -x+1, \cdots),
    \end{gathered}
    \label{eq:Sq1W}
\end{gather}
we derive the Wigner function corresponding to $\hat{\rho}_{\text{out}}$ as
\begin{align}
  & W_{\text{out}}(x_2, p_2) \nonumber \\
  & = \frac{2\pi}{P_{n=1}}
  \int_{-\infty} ^{\infty} \int_{-\infty} ^{\infty} dx_1 dp_1 W_{1}(x_1, p_1) W_{\text{G}} (x_1, x_2, p_1, p_2) \nonumber \\
  & = W_1 \left(e^{r_{\text{out}}} x_2, e^{-r_{\text{out}}} p_2\right).
\end{align}
This result represents a squeezed single-photon state $\hat{S}_2(r_{\text{out}})\ket{1}_2$ with a squeezing parameter $r_{\text{out}}$, which is written as 
\begin{align}
  e^{r_{\text{out}}} = e^{r_1 + r_2} \sqrt{\frac{2\sigma_{x,11} + 1}{2\sigma_{p,11} + 1}}.
  \label{eq:rout}
\end{align}
Here $\sigma_{x,11}$ and $\sigma_{p,11}$ are the $(1,1)$ components of $\sigma_{x}$ in Eq.~\eqref{eq:definitionsigma_x} and $\sigma_{p}$ in Eq.~\eqref{eq:definitionsigma_p}, respectively.
Whenever we choose three parameters $r_1, r_2$, and $T$ which realize the same $r_{\text{out}}$, the PNRD heralds the same pure squeezed single-photon state and only the single-photon detection probability $P_{n=1}$ in Eq.~\eqref{eq:P1} depends on these parameters.
The above discussion remains valid for GPS and PS ($r_2=0$) and thus the use of the PNRD avoids the trade-off relationship in the ideal lossless situation, while the use of an on/off detector causes the trade-off as discussed in the next section.

\subsection{PS and GPS with on/off detector\label{sec:PSandGPSwithon/offdetector}}
In this section, we quantitatively discuss the trade-off relationship in the case with the on/off detector by formulating the quality and generation rate of the heralded state.
First, we define the projection operator of the on/off detector on mode~1 as
\begin{align}
  \hat{\Pi}_{\text{off}} = \ket{0} \!{}_1 {}_1\!\bra{0}, \quad 
  \hat{\Pi}_{\text{on}} = \hat{I} - \ket{0} \!{}_1 {}_1\!\bra{0}, \label{eq:POVM}
\end{align}
where $\hat{I}$ is an identity operator.
Photon detection in mode~1 heralds a quantum state in mode~2 with the following density matrix $\hat{\rho}_{\text{out}}$:
\begin{align}
  \hat{\rho}_{\text{out}} = 
  \frac{\text{Tr}_1 \left( \hat{\Pi}_{\text{on}} \hat{\rho}_{\text{G}} \hat{\Pi}_{\text{on}}\right)}
  {P_{\text{on}}}
  = \frac{\text{Tr}_1 \hat{\rho}_{\text{G}} - {}_1\!\braket{0|\hat{\rho}_{\text{G}}|0}\!{}_1}{1-P_{\text{off}}}
  .
  \label{eq:rho}
\end{align}
Here zero-photon detection probability $P_{\text{off}}$ and photon detection probability $P_{\text{on}}$ are 
\begin{gather}
  \! \!
  P_{\text{off}} = \text{Tr} \left( \hat{\Pi}_{\text{off}} \hat{\rho}_{\text{G}} \right), \ 
  P_{\text{on}} = \text{Tr} \left( \hat{\Pi}_{\text{on}} \hat{\rho}_{\text{G}} \right) = 1-P_{\text{off}}.
  \label{eq:Pon}
\end{gather}
From the density matrix $\hat{\rho}_\text{out}$ in Eq.~\eqref{eq:rho}, we can calculate the corresponding Wigner function $W_{\text{out}}(x_2, p_2)$ as
\begin{align}
  \begin{split}
    &W_{\text{out}}(x_2, p_2) \\
    &= \frac{1}{1-P_{\text{off}}} \left[
    \int_{-\infty} ^{\infty} \int_{-\infty} ^{\infty} dx_1 dp_1 W_{\text{G}} (x_1, x_2, p_1, p_2) 
    \right.\\ 
    &\left.
    - 2\pi \int_{-\infty} ^{\infty} \int_{-\infty} ^{\infty} dx_1 dp_1 W_{0}(x_1, p_1) W_{\text{G}} (x_1, x_2, p_1, p_2)
    \right].
    \label{eq:W(x,p)}
  \end{split}
\end{align}
We here specifically examine $W_{\text{out}}(0,0)$ as a quality indicator.
This is because pure squeezed single-photon states have $W_{\text{out}}(0,0)=-1/\pi$, whereas $W_{\text{out}}(0,0)$ becomes larger as they get more contaminated by other heralded states or more deteriorated by losses.
Moreover, this indicator can be calculated analytically and compared with the experimental results easily.
In fact, it has been used for quality evaluation in previous experiments~\red{\cite{wenger_non-gaussian_2004,ourjoumtsev_generating_2006,neergaard-nielsen_generation_2006,wakui_photon_2007,dong_generation_2014,asavanant_generation_2017,takase_2022_generation}}.

\red{
  Note that we can also use fidelity between the heralded state and the target state as an alternative quality indicator.
  However, $W_\text{out}(0,0)$ and the fidelity are almost equivalent in our problem setting described later, so we only focus on $W_\text{out}(0,0)$ below.
  The detailed discussions are made in the Appendix.
}

From Eq.~\eqref{eq:W(x,p)}, $W_{\text{out}}(0,0)$ is calculated to be
\begin{gather}
  W_{\text{out}}(0,0) = \frac{1}{\pi}\frac{\frac{1}{2\sqrt{\sigma_{p, 11}\sigma_{x, 11}}} - P_{\text{off}}}{1-P_{\text{off}}},
  \label{eq:characteristics1}
\end{gather}
and $P_{\text{off}}$ is
\begin{align}
  P_{\text{off}} = \frac{2}{\sqrt{(2\sigma_{p, 11}+1)(2\sigma_{x, 11}+1)}}. \label{eq:Poff}
\end{align}
We here used Eqs.~\eqref{eq:VG}--\eqref{eq:Wout}, \eqref{eq:Sq1W} and the relationship $\sigma_x \sigma_p = I/4$.
Since the on/off detector mixes the desired squeezed single-photon state and other undesired states due to its incapability of distinguishing the number of photons, $W_{\text{out}}(0,0)$ in Eq.~\eqref{eq:characteristics1} is larger than $-1/\pi$ at any parameters.

Now we analyze the relationship between the quality indicator $W_{\text{out}}(0,0)$ and the photon detection probability $P_{\text{on}}$ for both PS and GPS, considering the practical scenario where we want to prepare the squeezed single-photon states \red{with the specific squeezing parameter $r_{\text{out}}$ by single-photon detection}.
In the following, we examine $W_{\text{out}}(0,0)$ and $P_{\text{on}}$ for various combinations of $r_1, r_2$, and $T$ where the same $\hat{S}_2(r_{\text{out}})\ket{1}_2$ is heralded by the single-photon detection.

\begin{figure}[tbhp]
  \begin{subfigure}[t]{\linewidth}
    \centering
    \subcaption{
      \label{fig:NegativityVSCountRate}
      }
      \includegraphics{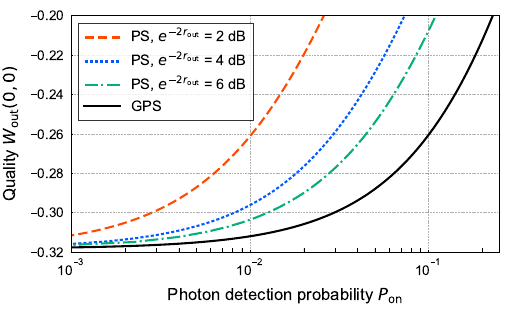}
  \end{subfigure}
  \\
  \begin{subfigure}[t]{\linewidth}
    \centering
    \subcaption{
      \label{fig:LossEffect2dB}
      }
    \includegraphics{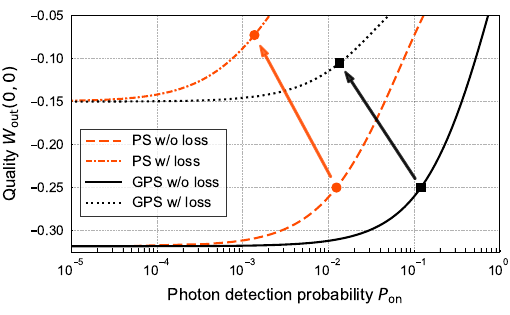}
  \end{subfigure}
  \caption{Numerical results of trade-offs.
  (a) Trade-off comparison between PS and GPS \red{for squeezed single-photon states with specific squeezing levels.}
  As for GPS, we plot the best trade-off obtained by optimizing the three parameters $r_1, r_2$, and $T$, and this trade-off is independent of the output squeezing parameter $r_{\text{out}}$ \red{ of the squeezed single-photon states which are heralded by single-photon detection.}
  (b) Impact of losses on the trade-off improvement.
  The black solid line and orange dashed line are trade-off lines in the case of $e^{-2r_\text{out}}=\SI{2}{\deci\bel}$ without losses, which are the same as those in (a).
  Addition of losses without changing other conditions shifts these two lines toward the black dotted line and the orange dot-dashed line, respectively.
  We include signal channel loss of 25\% and trigger channel loss of 90\% (equivalent to the loss values of our experimental setup, described later) in our simulations with the losses.
  The circle and square symbols indicate the example of how the losses displace the points that have the same quality $W_{\text{out}}(0,0)=-0.25$ for PS and GPS in lossless situations as indicated by arrows. 
  }
\end{figure}

First, we discuss the case of using PS, which has only two parameters $r_1$ and $T$ ($r_2=0$).
Once we choose the transmissivity $T$, the input squeezing parameter $r_1$ is automatically determined by Eq.~\eqref{eq:rout}, and both $W_{\text{out}}(0,0)$ and $P_{\text{on}}$ are obtained by Eqs.~\eqref{eq:Pon}, \eqref{eq:characteristics1}, and \eqref{eq:Poff}.
Then, the relationship between $W_{\text{out}}(0,0)$ and $P_{\text{on}}$ is determined by changing $T$ as a parameter.
Figure~\ref{fig:NegativityVSCountRate} shows the numerical results of this relationship for different $r_{\text{out}}$.
We see a better trade-off between $W_{\text{out}}(0,0)$ and $P_{\text{on}}$ for higher $r_{\text{out}}$. 

Next, considering GPS, where three parameters $r_1$, $r_2$, and $T$ are tunable for the best trade-off as shown below, we first reformat $W_{\text{out}}(0,0)$ in Eq.~\eqref{eq:characteristics1} to
\begin{gather}
  W_{\text{out}}(0,0) = \frac{1}{\pi}\frac{\frac{P_{\text{off}}}{\sqrt{P_{\text{off}}^2 - 2P_{\text{off}}(s+s^{-1})+4}}-P_{\text{off}}}{1-P_{\text{off}}}
  \label{eq:UsingS}
\end{gather}
by introducing a new parameter $s$:
\begin{align}
  s = e^{r_{\text{out}}-r_1-r_2}>0. \label{eq:s}
\end{align}
Here we suppose that the output squeezing parameter and photon detection probability are given by $r_{\text{out}}=r_c$ and $P_{\text{on}}=P_c$, respectively.
From Eqs.~\eqref{eq:rout}, \eqref{eq:characteristics1}, \eqref{eq:Poff} and $0<T<1$, $r_1$ and $r_2$ should satisfy the following conditions:
\begin{gather}
    \cosh{r_1} \cosh{r_2} = \frac{\cosh{r_c}}{1-P_c}, \\
    \left(r_1 - r_c\right) \left(r_2 - r_c\right) < 0,
\end{gather}
and this derives the possible range of $s$ as 
\begin{align}
  \begin{dcases}
    0 < s < \exp{\left[\text{arcosh}\left(\frac{1}{1-P_c}\right)\right]}
    \ & \text{if} \quad r_c > 0 \\
    s > 0
    \ & \text{if} \quad r_c = 0 \\
    s > \exp{\left[-\text{arcosh}\left(\frac{1}{1-P_c}\right)\right]}
    \ & \text{if} \quad r_c < 0 \\
  \end{dcases}.
\end{align}
We finally minimize $W_{\text{out}}(0,0)$ in Eq.~\eqref{eq:UsingS} by changing the parameter $s$ within the above range.
Regardless of the values of $r_c$ and $P_c$, we can minimize $W_{\text{out}}(0,0)$ at $s=1$, where the specific combinations of $r_1$, $r_2$, and $T$ are determined by Eqs.~\eqref{eq:rout} and \eqref{eq:s}.
At such minimum point, Eq.~\eqref{eq:UsingS} becomes
\begin{align}
  W_{\text{out}}(0,0) = -\frac{1}{\pi}\frac{P_{\text{off}}}{2-P_{\text{off}}}=-\frac{1}{\pi}\frac{1-P_{\text{on}}}{1+P_{\text{on}}},
\end{align}
which is the best trade-off in GPS.
This trade-off relationship is independent of the output squeezing parameter $r_{\text{out}}$, unlike PS.
We plot the numerical result of this best trade-off in Fig.~\ref{fig:NegativityVSCountRate} and see that GPS (black solid line) achieves a better trade-off than PS (the other three lines).
Furthermore, this improvement is theoretically valid at any $r_{\text{out}}$.
GPS with three adjustable parameters $r_1$, $r_2$, and $T$ can reach $s=1$ as described above, while PS with only two parameters $r_1$ and $T$ ($r_2=0$) cannot.
Hence, GPS achieves smaller $W_{\text{out}}(0,0)$ for the given $P_{\text{on}}$ at any $r_{\text{out}}$; in other words, GPS increases $P_{\text{on}}$ for the same $W_{\text{out}}(0,0)$ at any $r_{\text{out}}$.
Particularly in the weak output squeezing regimes, such as \SI{2}{\deci\bel}, the photon detection probability $P_{\text{on}}$ becomes approximately 10 times larger in GPS for the same quality $W_{\text{out}}(0,0)$.

So far we have discussed the ideal situation without optical losses such as propagation losses and inefficiencies of the on/off detector.
Here we briefly mention the practical situation with these losses by using Fig.~\ref{fig:LossEffect2dB} as an example.
In Fig.~\ref{fig:LossEffect2dB}, the black solid line and orange dashed line are trade-off lines for $e^{-2r_{\text{out}}}=\SI{2}{\deci\bel}$, which are the same as those plotted in Fig.~\ref{fig:NegativityVSCountRate}.
In contrast, the other two lines are obtained by using the same conditions as the already mentioned two lines but including a certain amount of losses. 
The circle and square symbols in this figure indicate an example of how the losses displace the points that have the same quality $W_{\text{out}}(0,0)=-0.25$ for PS and GPS in lossless situations.
By focusing on these symbols, we can qualitatively discuss the impact of the losses on $P_{\text{on}}$, $W_{\text{out}}(0,0)$, and consequently the trade-off line, as follows.
In the typical experiment where we generate high-quality states with $W_{\text{out}}(0,0)<0$, the multiphoton detection probability is small compared with the single-photon detection probability.
Hence, $P_{\text{on}}$ decreases almost linearly with respect to the trigger channel loss in the photon detection side, and the trade-off lines of both PS and GPS shrink uniformly by the same scale [by almost a factor of 10 in Fig.~\ref{fig:LossEffect2dB}] along the horizontal axis.
As for $W_{\text{out}}(0,0)$, their degradations are mainly determined by the signal loss.
Additionally taking into consideration that the squeezed single-photon states, which are dominant in the heralded states, are degraded in the same way in both PS and GPS, $W_{\text{out}}(0,0)$ of the heralded states by both PS and GPS increase by approximately the same degree.
Thereby the trade-off lines of both PS and GPS are shifted almost similarly along both the horizontal and vertical axes.
As a result, the generation rate is increased by GPS as well even when the effect of losses is included, and the degree of this increase does not change.
Note that the losses slightly change the combinations of the parameters for the best trade-off in GPS, but at least the difference is expected to be small under the amount of losses we consider in this work.
Analytically deriving these combinations is left for future work.

\section{Experimental setup \& Data acquisition\label{sec:Experimentalsetup}}
\begin{figure*}[t]
  \centering
  \includegraphics[width=\linewidth]{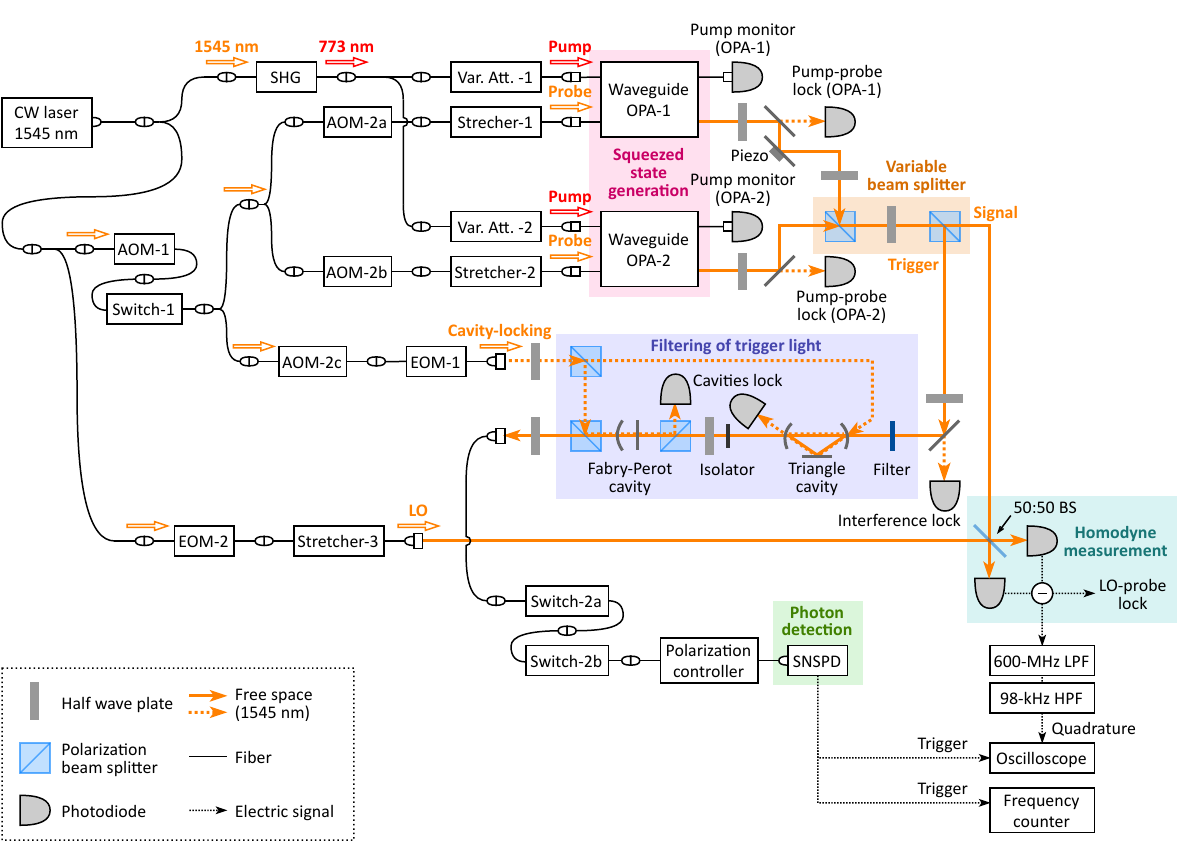}
  \caption{
    Experimental setup.
    CW, continuous wave;
    SHG, second harmonic generator;
    AOM, acousto-optic modulator;
    Var. Att., variable attenuator;
    OPA, optical parametric amplifier;
    EOM, electro-optic modulator;
    SNSPD, superconducting nanostrip single-photon detector;
    LO, local oscillator;
    BS, beam splitter;
    LPF, low-pass filter;
    HPF, high-pass filter
    .
    \label{fig:setup}}
\end{figure*}

To experimentally prove our theoretical conclusion in Sec.~\ref{sec:TheoryNumericalresults}, we constructed the setup as shown in Fig.~\ref{fig:setup}, which was applicable to both PS and GPS with an on/off detector.
\red{Although our discussion in Sec.~\ref{sec:TheoryNumericalresults} is valid for both experiments using the pulsed laser and the continuous-wave laser, our demonstration was conducted with the latter one.}
We use a continuous-wave laser with the center wavelength at \SI{1545}{\nano\meter}.
Its second-harmonic light at wavelength of \SI{773}{\nano\meter} is divided and injected into two fiber-coupled waveguide optical parametric amplifier (OPA) modules~\cite{kashiwazaki_fabrication_2021} as the pump light to generate squeezed states.
Here we have variable attenuators (\mbox{Var.~Att.-1} and \mbox{Var.~Att.-2}) before each OPA to change the input pump power that determines the squeezing level.

Then, the output squeezed states are interfered with each other in free space by a variable beam splitter (VBS).
This VBS consists of two polarization beam splitters (PBSs) and one half-wave plate (HWP), so its transmissivity can be set arbitrarily by changing the rotation angle of the HWP manually.

Next, each output of this VBS goes to a homodyne measurement and a photon detection system, respectively.
The light to the homodyne measurement system corresponds to the signal mode, and its quadrature amplitude is measured with the homodyne detector.
Then, the detector's output is acquired with the oscilloscope after electrical filtering.
The other output of the VBS, corresponding to the trigger mode, first goes through the spectral filtering system.
This filtering restricts the frequency region of the detected photons, and consequently limits that of the heralded squeezed single-photon states within our homodyne measurement bandwidth, $\sim$\SI{200}{\mega\hertz}.
Our filtering system has three parts: a dielectric multi-layer filter with half width at half maximum (HWHM) of \SI{0.5}{\nano\meter}, a triangle cavity with HWHM of \SI{2.8}{\mega\hertz} and free spectral range (FSR) of \SI{4.8}{\giga\hertz}, and a Fabry-Perot cavity with HWHM of \SI{160}{\mega\hertz} and FSR of \SI{162}{\giga\hertz}.
This combination ensures that only the trigger light within HWHM of \SI{2.8}{\mega\hertz} around the center wavelength of the laser transmits the filtering system.
Then, this filtered trigger light goes to a superconducting nanostrip single-photon detector (SNSPD)~\cite{miki_stable_2017} after the polarization controller.
The detection efficiency of this SNSPD is $\sim$65\%, and the dark count rate is $\SI{\sim25}{cps}$ (not corrected by the ratio of the measurement period, which is described later) at the typical bias current applied to the nanostrip.

In Fig.~\ref{fig:setup}, there is other miscellaneous apparatus required for optical phase control.
To generate and observe phase-sensitive quantum states, we have to fix the relative phases between the pump light and the local oscillator (LO) light of the homodyne measurement.
We alternately repeat the phase control period and measurement period at \SI{2}{\kilo\hertz}.
During the control period, we input additional classical light at \SI{1545}{\nano\meter}, called \textit{probe} light, into each OPA.
The frequencies of this probe light are shifted by acousto-optic modulators (AOM-1, AOM-2a, and AOM-2b) from the original frequency for the purpose of phase locking with a coherent control sideband method~\cite{vahlbruch_coherent_2006,chelkowski_coherent_2007}.
We lock the relative phases with fiber stretchers and a piezo-mounted mirror so that we measure the antisqueezed quadrature of the OPA-1's output squeezed state by default.
This probe light is blocked by optical switches (Switch-2a and Switch-2b) before the SNSPD.
We also input additional classical light into the filtering cavities and fix each cavity length by the Pound-Drever-Hall method~\cite{black_introduction_2001} so that its resonant frequency matches the incoming light frequency.

During the measurement period, we block the probe light and cavity-locking light by turning off all the AOMs and Switch-1.
Then, we shift the LO phase using an electro-optic modulator (EOM-2) from the locked phase to the target phase for the measurement.
Meanwhile, we turn on Switch-2a and Switch-2b to allow the trigger photon to reach the SNSPD.

Using the above setup, we performed the experiments of PS and GPS as follows.
We first adjusted the pump power and the transmissivity of the VBS to the target conditions ($r_1$, $r_2$, and $T$).
Then, we obtained the photon detection rate, which corresponds to $P_{\text{on}}$, by measuring the photon detection signal from the SNSPD with a frequency counter.
Meanwhile, we acquired the homodyne detector's output using the photon detection signal as a trigger to evaluate the quality of the heralded states.
We measured \num{5000} time waveforms of quadratures at 12 phases in \SI{15}{\degree} increments.
Then, we calculated the quadrature amplitude of a wave-packet mode which was automatically determined by the experimental setup~\cite{yoshikawa_purification_2017}.
Its temporal mode function is 
\begin{align}
  f(t) \propto
  \begin{dcases}
    e^{2\pi\gamma_1 (t-t_0)} - e^{2\pi\gamma_2 (t-t_0)} \quad & \text{if} \quad t\leq t_0 \\
    0 & \text{if} \quad t>t_0 \\
  \end{dcases}
\end{align}
without normalization.
Theoretically, the parameters $\gamma_1$ and $\gamma_2$ are determined by the HWHM of two filter cavities before the SNSPD, and $t_0$ is the temporal parameter that corresponds to the photon detection timing.
In the data analysis, they were optimized based on the experimental results ($\gamma_1=\SI{2.8}{\mega\hertz}$ and $\gamma_2=\SI{80}{\mega\hertz}$).
Finally, we reconstructed the Wigner functions from these quadratures using the maximum likelihood estimation method~\cite{lvovsky_iterative_2004}.
\red{Note that the wave-packet mode can be modified by changing the filtering system.
Although it changes the generation rate for PS and GPS by the same factor, it does not change the essential performance and formulation of these schemes.}

\section{Experimental Results \& Discussion\label{sec:ExperimentalResultsDiscussion}}
In this section, we show the experimental results of GPS performed with the setup of Fig.~\ref{fig:setup}.

\subsection{Comparison between theoretical and experimental GPS\label{sec:GenerationexamplesusingGPSmethod}}
First, we validate our theoretical model of GPS in Sec.~\ref{sec:TheoryNumericalresults} by comparing experimentally generated states with the corresponding simulation results.
We used three combinations of $r_1$ and $r_2$ to observe how the heralded states and photon detection rates depend on the BS transmissivity $T$.

\begin{figure*}[htbp]
  \centering
  \begin{subfigure}{\linewidth}
    \subcaption{
      \label{fig:SimVSExp_3dBT_2}
      }
    \includegraphics{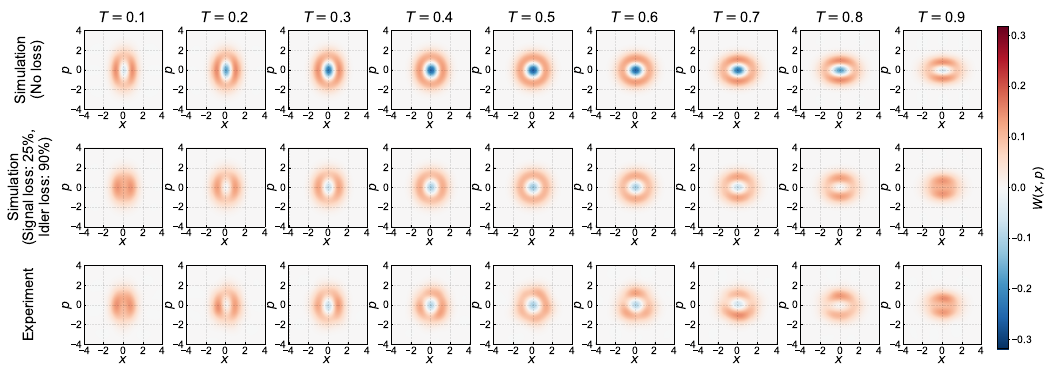}
  \end{subfigure}
  \\
  \begin{subfigure}{\linewidth}
    \subcaption{
      \label{fig:SimVSExp_2dBT_2}
      }
    \includegraphics{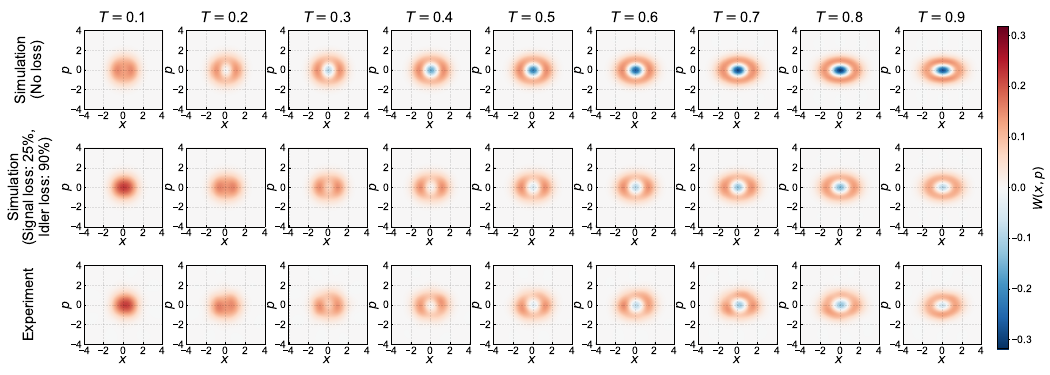}
  \end{subfigure}
  \\
  \begin{subfigure}{\linewidth}
    \subcaption{
      \label{fig:SimVSExp_4dBT_2}
      }
    \includegraphics{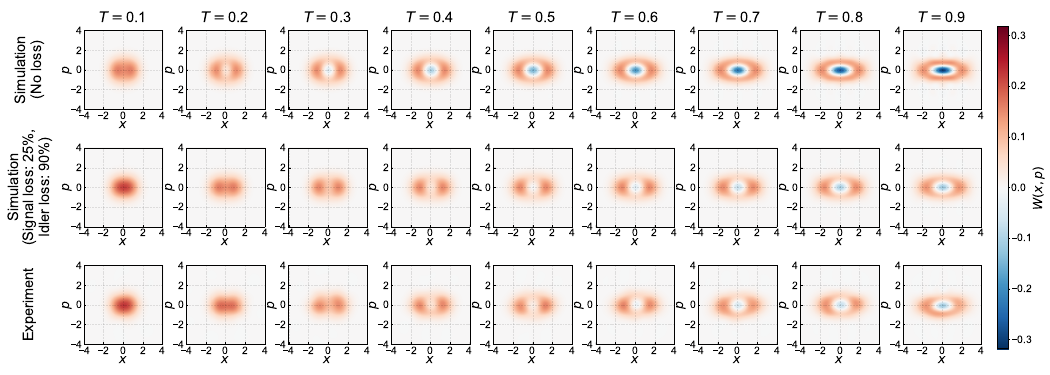}
  \end{subfigure}
  \caption{
    Wigner functions of the quantum states generated by various combinations of three parameters $r_1$, $r_2$, and $T$.
    Three subfigures correspond to the following combinations of input squeezing levels: 
    (a) $e^{-2r_1} = \SI{3.0}{\deci\bel}, \  e^{-2r_2} = \SI{-3.0}{\deci\bel}$;
    (b) $e^{-2r_1} = \SI{2.8}{\deci\bel}, \  e^{-2r_2} = \SI{-0.78}{\deci\bel}$;
    (c) $e^{-2r_1} = \SI{4.7}{\deci\bel}, \  e^{-2r_2} = \SI{-0.67}{\deci\bel}$.
    \label{fig:WignerFunctions}
  }
\end{figure*}

Figure~\ref{fig:WignerFunctions} is the comparison between the simulation and experimental results of the Wigner functions.
Figures~\ref{fig:SimVSExp_3dBT_2}--\ref{fig:SimVSExp_4dBT_2} are obtained at the following combinations of two input squeezing levels $e^{-2r_1}$ and $e^{-2r_2}$: (a) \SI{3}{\deci\bel} and \SI{-3}{\deci\bel} (symmetric condition, $|r_1|=|r_2|$), (b) \SI{2.8}{\deci\bel} and \SI{-0.78}{\deci\bel} (weakly asymmetric condition), and (c) \SI{4.7}{\deci\bel} and \SI{-0.67}{\deci\bel} (strongly asymmetric condition).
Within each subfigure, the different columns represent the results at various BS transmissivities with 0.1 intervals.
We plot the experimental results in the lower rows, and the upper and central rows are the corresponding simulation results without and with the losses inherent in our experimental setup, respectively.
In our simulations, we include the signal channel loss of 25\% and the trigger channel loss of 90\%.
The breakdowns of the signal losses are shown in Table~\ref{table:signal}.
Among them, the fake photon detection triggers are divided into two types: dark counts of the SNSPD and undesired detections of slight reflection of the LO light from the surface of photodiodes in the homodyne detector.
The former counts do not change among different experiments, $\SI{\sim25}{cps}$.
In contrast, the latter counts depend on experimental conditions such as BS transmissivity and also fluctuate over time during the measurement.
Hence, the typical value is written in the table.
On the other hand, the trigger channel loss involves the propagation loss between the VBS and the SNSPD, $\sim$80\%, and the detection efficiency of the SNSPD, $\sim$65\%.

\begin{table}[hbtp]
  \caption{Loss budget of signal path}
  \label{table:signal}
  \centering
  \begin{ruledtabular}
  \begin{tabular}{lr}
    Item  &  Typical value\\
    \hline
    Waveguide OPA  & 9\%\\
    Propagation (total)  & 5\%\\
    Spatial mode mismatch & 4\% \\
    Inefficiency of photodiodes  &  1\%\\
    Circuit noise of homodyne detector & 1\%\\
    \SI{98}{\kilo\hertz} cutoff HPF & 3\%\\
    Fake photon detection triggers of SNSPD & $\sim$5\%\\
    \hline
    Total & $\sim$25\%\\
  \end{tabular}  
  \end{ruledtabular}
\end{table}

\begin{figure*}[htbp]
  \begin{minipage}[t]{0.49\linewidth}
    \centering
    \begin{subfigure}{\linewidth}
      \subcaption{
        \label{fig:NegativityMin_3dBT_2}
        }
      \includegraphics{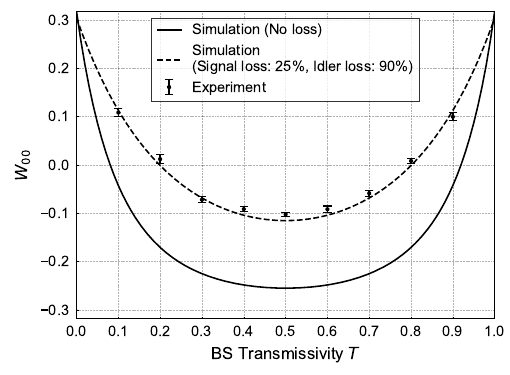}
    \end{subfigure}
    \begin{subfigure}{\linewidth}
      \subcaption{
        \label{fig:NegativityMin_2dBT_2}
        }
      \includegraphics{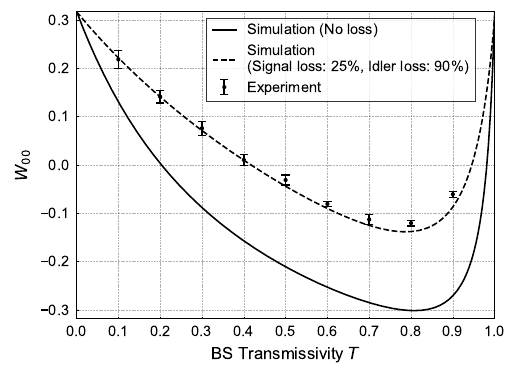}
    \end{subfigure}
    \begin{subfigure}{\linewidth}
      \subcaption{
        \label{fig:NegativityMin_4dBT_2}
        }
      \includegraphics{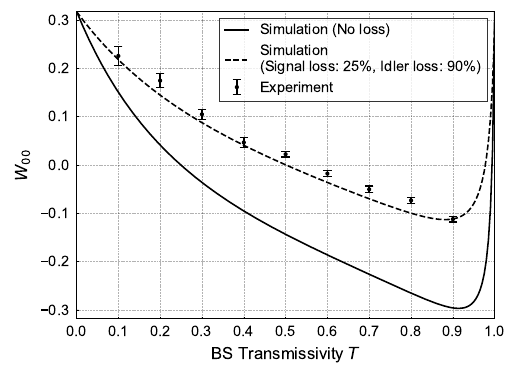}
    \end{subfigure}
  \end{minipage}
  \begin{minipage}[t]{0.49\linewidth}
    \centering
    \begin{subfigure}{\linewidth}
      \subcaption{
        \label{fig:CountRateZoom_3dBT_2}
        }
      \includegraphics{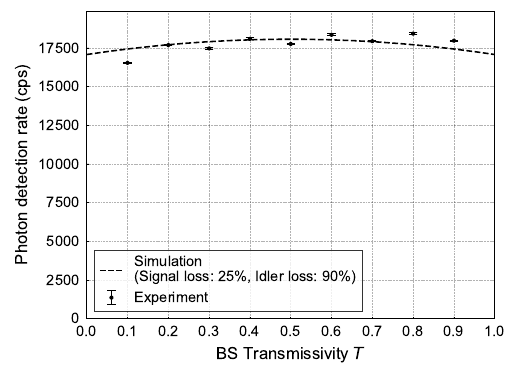}
    \end{subfigure}
    \begin{subfigure}{\linewidth}
      \subcaption{
        \label{fig:CountRateZoom_2dBT_2}
        }
      \includegraphics{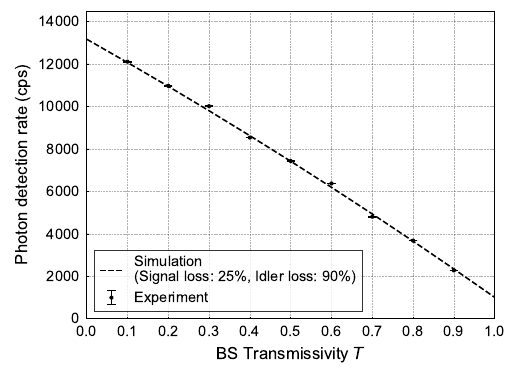}
    \end{subfigure}
    \begin{subfigure}{\linewidth}
      \subcaption{
        \label{fig:CountRateZoom_4dBT_2}
        }
      \includegraphics{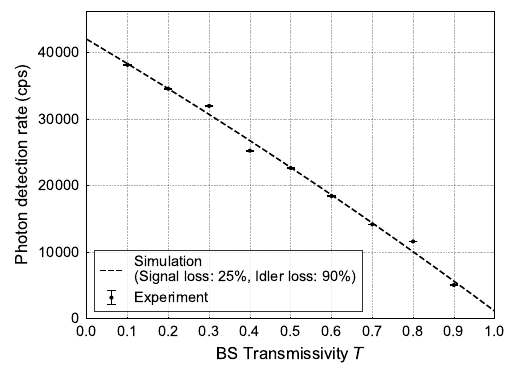}
    \end{subfigure}
  \end{minipage}
  \caption{
      Quantitative evaluation of quantum states shown in Fig.~\ref{fig:WignerFunctions}.
      (a)--(c) Quality $W_{00}$ of quantum states corresponding to Figs.~\ref{fig:SimVSExp_3dBT_2}--\ref{fig:SimVSExp_4dBT_2}, respectively.
      (d)--(f) Photon detection rates corresponding to Figs.~\ref{fig:SimVSExp_3dBT_2}--\ref{fig:SimVSExp_4dBT_2}, respectively.
      We plot simulation results of photon detection probabilities $P_{\text{on}}$ by multiplying appropriate scaling constants $C$, which we obtain through the following fitting.
      In each subfigure, we obtained nine values of the ratio between the experimental detection rate and the corresponding $P_{\text{on}}$, and then used their average for $C$.
      The following is the values of $C$ in each subfigure: 
      (d)~\num[exponent-product = \times]{1.47e6},
      (e)~\num[exponent-product = \times]{1.31e6}, and 
      (f)~\num[exponent-product = \times]{1.43e6}.
      Here the differences among these scaling constants are attributed to different trigger channel loss values in different measurements.
      All the error bars indicate the standard errors.
    \label{fig:NegativityPhotonDetectionRate}
    }
\end{figure*}

Figure~\ref{fig:WignerFunctions} shows that our experimental Wigner functions visually agree well with the simulations.
Moreover, this figure clearly illustrates the general characteristics of GPS with the on/off detector.
In the simulations without losses in Fig.~\ref{fig:SimVSExp_3dBT_2}, $W(0,0)$ is minimized at $T=0.5$ and the negative values of $W(0,0)$ gradually fade away as the transmissivity is set closer to $T=0 \ \text{or} \ 1$.
The squeezing degree and direction of the Wigner function of the generated states also change depending on $T$.
In Figs.~\ref{fig:SimVSExp_2dBT_2} and \ref{fig:SimVSExp_4dBT_2}, we can see a similar trend, but with different optimal values of $T$ which minimize $W(0,0)$ depending on the combination of $r_1$ and $r_2$.
This trend is observed also in the simulations with losses and experiments. 
It is a distinctive feature of GPS compared to conventional PS, where $W(0,0)$ changes monotonically with respect to $T$ and is not minimized within $0<T<1$.
Such difference results from one additional degree of freedom $r_2$ in GPS, which also contributes to the trade-off improvement discussed in Sec.~\ref{sec:TheoryNumericalresults}.

For a more quantitative assessment, we graphically show how the values of $W(0,0)$ as a quality indicator depend on $T$ in the left column of Fig.~\ref{fig:NegativityPhotonDetectionRate}.
Here the plotted values $W_{00}$ are defined as follows:
\begin{align}
  W_{00} = 
  \begin{dcases}
    \min W(x,p) \ &\text{if} \quad W(0,0)<0\\
    W(0,0) \ &\text{if} \quad W(0,0)\geq 0\\
  \end{dcases}
  ,
\end{align}
neglecting the small displacements of the minimum points of the experimental Wigner functions, which are caused by some experimental reasons.
Their errors plotted in Fig.~\ref{fig:NegativityPhotonDetectionRate} were calculated by the Fisher information matrix~\cite{rehacek_tomography_2008}.
The behavior of $W_{00}$ in relation to $T$ is well matched between the simulations and experiments, for example, in terms of where they reach their minimum.
This consistency ensures the validity of our theoretical model.
Note that the slight deviation between the experimental and simulation results of $W_{00}$ is mainly attributed to the fluctuation of the signal loss from 25\%.

As for photon detection rates, we show how they depend on $T$ in the right column of Fig.~\ref{fig:NegativityPhotonDetectionRate}.
We plot their averages for \SI{1}{\minute} with their standard errors.
These are raw values that are not corrected by the duty cycle of the measurement period, 13\%.
We also plot the simulation results by multiplying the photon detection probabilities $P_{\text{on}}$, defined by Eqs.~\eqref{eq:Pon} and \eqref{eq:Poff}, by appropriate scaling constants (details in the caption of Fig.~\ref{fig:NegativityPhotonDetectionRate}).
We first find that the experimental rates closely follow the simulation curves.
Then, we see a drastic dependence of photon detection rates on $T$ in Figs.~\ref{fig:CountRateZoom_2dBT_2} and \ref{fig:CountRateZoom_4dBT_2}, where the input squeezing levels are asymmetric.
In such cases, as the larger portion of the squeezed state with the higher squeezing level of the two inputs reaches the on/off detector, the photon detection rate becomes higher.
Additionally, the photon detection rates do not converge to zero even at $T\rightarrow1$, which is a different characteristic from PS.

Thus by comparing the experimental results with the simulation results, we validated our theoretical model of GPS.
This consequently ensures our discussion of the trade-off improvement in Sec.~\ref{sec:TheoryNumericalresults}, and then it is experimentally confirmed in the next section.

\subsection{Improvement of trade-off by GPS\label{sec:ComparisonbetweenPSandGPS}}
\begin{figure*}[!bt]
  \begin{subfigure}[t]{0.487\linewidth}
    \centering
    \subcaption{
      \label{fig:PSvsGPS_2dB_2}
      }
      \includegraphics[trim=0 0 2.5mm 0, clip]{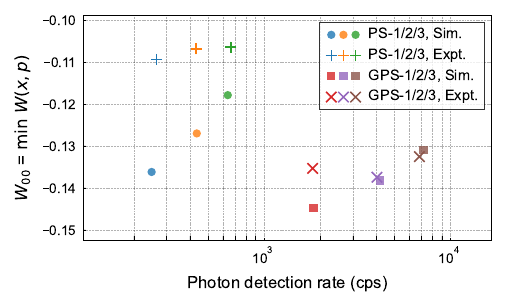}
  \end{subfigure}
  \begin{subfigure}[t]{0.25\linewidth}
    \centering
    \subcaption{
      \label{fig:Wigner_0830_1PS2_95_v1}
    }
    \includegraphics[trim=1mm 0 1mm 1.5mm, clip]{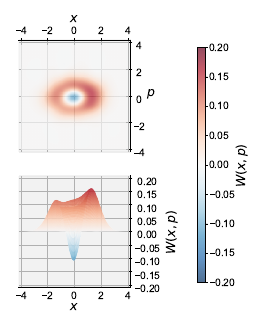}
  \end{subfigure}
  \begin{subfigure}[t]{0.25\linewidth}
    \centering
    \subcaption{
      \label{fig:Wigner_0830_1GPS2_78_v1}
      }
      \includegraphics[trim=1mm 0 1mm 1.5mm, clip]{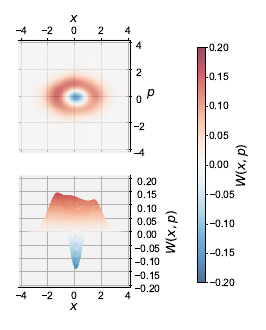}
    \end{subfigure}
    \\
    \begin{subfigure}[t]{0.487\linewidth}
      \centering
      \subcaption{
        \label{fig:PSvsGPS_4dB_2}
        }
        \includegraphics[trim=0 0 2.5mm 0, clip]{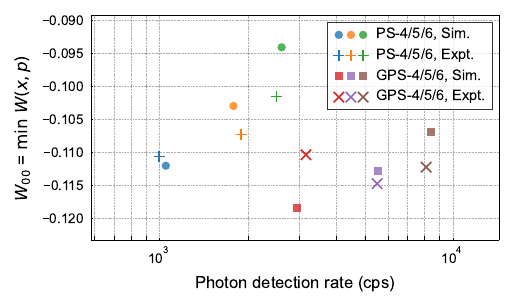}
      \end{subfigure}
      \begin{subfigure}[t]{0.25\linewidth}
        \centering
        \subcaption{
          \label{fig:Wigner_0830_1PS4_95_v1}
          }
          \includegraphics[trim=1mm 0 1mm 1.5mm, clip]{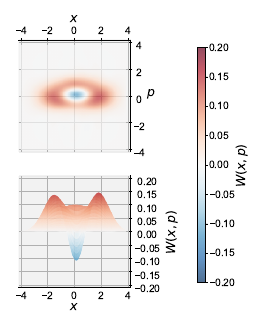}
        \end{subfigure}
        \begin{subfigure}[t]{0.25\linewidth}
          \centering
          \subcaption{
            \label{fig:Wigner_0830_1GPS4_88_v1}
            }
            \includegraphics[trim=1mm 0 1mm 1.5mm, clip]{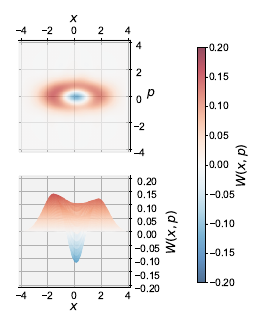}
          \end{subfigure}
          \caption{
            \red{Experimental comparison between PS and GPS for squeezed single-photon states with the same squeezing levels.}
            (a) and (d) Experimental and simulation results of trade-off of 2-dB- and 4-dB-squeezed single-photon states, respectively.
            For scaling the simulation results of photon detection probabilities $P_{\text{on}}$, we use the following values of $C$:
            (a)~\num[exponent-product = \times]{1.45e6} and
            (d)~\num[exponent-product = \times]{1.30e6}. 
            (b), (c), (e), and (f) Representative Wigner functions of quantum states generated by the following conditions in the order: PS-2, GPS-2, PS-5, and GPS-5. 
            Refer to Tables~\ref{table:PSvsGPScondition2dB} and \ref{table:PSvsGPScondition4dB} for the experimental conditions and detailed results.
            \label{fig:TradeoffResults}
            }
  \end{figure*}

Finally, we experimentally prove our theoretical derivation that GPS can improve the trade-off between the quality and photon detection rate compared with PS.
We generated 2-dB- and 4-dB-squeezed single-photon states using different experimental parameters $r_1$, $r_2$, and $T$ by both PS and GPS, analyzed the data using the same method as Sec.~\ref{sec:GenerationexamplesusingGPSmethod}, and then compared the results between PS and GPS.

Figures~\ref{fig:PSvsGPS_2dB_2} and \ref{fig:PSvsGPS_4dB_2} show the trade-off between the qualities and photon detection rates in the cases of \SI{2}{\deci\bel} and \SI{4}{\deci\bel}, respectively.
In each of these subfigures, we plot three results for each of PS and GPS, which were obtained by different combinations of $r_1$, $r_2$, and $T$.
We also plot the corresponding simulation results with losses by scaling the photon detection probabilities similarly to Figs.~\ref{fig:CountRateZoom_3dBT_2}--\ref{fig:CountRateZoom_4dBT_2}.
Tables~\ref{table:PSvsGPScondition2dB} and \ref{table:PSvsGPScondition4dB} list the experimental conditions and detailed results.
In GPS, we used the parameters $r_1$, $r_2$, and $T$ approximately satisfying the constraint for the best trade-off.
As mentioned in Sec.~\ref{sec:TheoryNumericalresults}, their specific values are determined by Eqs.~\eqref{eq:rout} and \eqref{eq:s} ($s=1$ at the best trade-off).
  
We discuss the trade-off results as follows.
In Figs.~\ref{fig:PSvsGPS_2dB_2} and \ref{fig:PSvsGPS_4dB_2}, the experimentally obtained photon detection rates (horizontal axis) agree well with the simulations, and all the experimental values in GPS surpass those in PS.
As for $W_{00}$ (vertical axis), our experiments exhibit a similar trend to the simulations, and all the experimental values in GPS are either equivalent to or better than those in PS.
The reason for differences in the qualities $W_{00}$ between the experiments and simulations is that 
the negative value $W_{00}$ is sensitive to the signal loss values, which depend on the alignment in each measurement.
In particular, relatively large discrepancies in PS-1, 2, and 3 are attributed to the fake photon detection triggers.
In these experimental conditions, the photon detection rates and fake count rates are several \SI{100}{cps} and several \SI{10}{cps}, respectively, which corresponds to almost 10\% loss.
This is higher than 5\% loss in Table~\ref{table:signal}, which is used in our simulation as shown in Table~\ref{table:signal}.
As a whole, as Figs.~\ref{fig:PSvsGPS_2dB_2} and \ref{fig:PSvsGPS_4dB_2} show, the trade-off relationship becomes better by adopting GPS in both cases of \SI{2}{\deci\bel} and \SI{4}{\deci\bel}.
At \SI{2}{\deci\bel} in particular, the photon detection rate in GPS exceeds that in PS by over 10 times.

Here let us confirm that the generated states are close to each other in six different experimental conditions for \SI{2}{\deci\bel} and \SI{4}{\deci\bel}, respectively.
Figure~\ref{fig:TradeoffResults} shows representative Wigner functions.
We here see that PS-2 and GPS-2 in Figs.~\ref{fig:Wigner_0830_1PS2_95_v1} and \ref{fig:Wigner_0830_1GPS2_78_v1} as well as PS-5 and GPS-5 in Figs.~\ref{fig:Wigner_0830_1PS4_95_v1} and \ref{fig:Wigner_0830_1GPS4_88_v1} have almost equivalent squeezing levels, respectively.
In addition, we quantitatively evaluated the squeezing levels \red{$e^{-2r_{\text{est}}}$} of the actually generated squeezed single-photon states by fitting their Wigner functions with those of the pure squeezed single-photon states.
We list the obtained values as the \red{estimated output squeezing levels \red{$e^{-2r_{\text{est}}}$}} in Tables~\ref{table:PSvsGPScondition2dB} and \ref{table:PSvsGPScondition4dB}, and their discrepancies among different experimental conditions are within \SI{\pm0.10}{\deci\bel}.
Note that the differences between these values and the target squeezing levels (\SI{2}{\deci\bel} and \SI{4}{\deci\bel}) mainly arise from the fact that actually generated states are deteriorated by the losses.

Thus we experimentally confirmed our theoretical prediction of the trade-off improvement by GPS.

\sisetup{separate-uncertainty}
\begin{table*}[!t]
  \begin{minipage}[t]{\linewidth}
    \begin{threeparttable}
      \caption{Experimental conditions and results of 2-dB-squeezed single-photon states}
      \label{table:PSvsGPScondition2dB}
      \centering
      \begin{ruledtabular}
    \begin{tabular}{lcccccc}
      & \multicolumn{2}{c}{\multirow{2}{*}{Input squeezing levels}}
      & \multirow{2}{*}{Transmissivity}
      & \red{Estimated output}
      & \multirow{2}{*}{Quality\tnote{*}}
      & \multirow{2}{*}{Photon detection rate\tnote{*}}
      \\
      & \multicolumn{2}{c}{} & & \red{squeezing level} & &
      \\
      & $e^{-2r_1}$ (dB) & $e^{-2r_2}$ (dB)
      & $T$
      & \red{$e^{-2r_{\text{est}}}$} (dB)
      & $W_{00} = \min W(x,p)$
      & (cps)\\
      \hline
      PS-1 & 2.06 & 0 & 0.97 & 1.62 & $-0.109 \pm 0.010$ & \num{2.65(2)e2} \\
      PS-2 & 2.11 & 0 & 0.95 & 1.58 & $-0.107 \pm 0.009$ & \num{4.32(3)e2} \\
      PS-3 & 2.16 & 0 & 0.93 & 1.70 & $-0.106 \pm 0.009$ & \num{6.63(4)e2} \\
      GPS-1 & 2.40 & $-0.39$ & 0.86 & 1.63 & $-0.135 \pm 0.008$ & \num{1.82(1)e3} \\
      GPS-2 & 2.80 & $-0.78$ & 0.79 & 1.63 & $-0.137 \pm 0.007$ & \num{4.03(2)e3} \\
      GPS-3 & 3.20 & $-1.14$ & 0.74 & 1.58 & $-0.133 \pm 0.005$ & \num{6.82(2)e3} \\
    \end{tabular}  
  \end{ruledtabular}
  \footnotesize
  \begin{tablenotes}
    \item[*] Shown with the standard errors.
  \end{tablenotes}
\end{threeparttable}
  \end{minipage}
  \\
  \begin{minipage}[t]{\linewidth}
    \begin{threeparttable}
    \caption{Experimental conditions and results of 4-dB-squeezed single-photon states}
    \label{table:PSvsGPScondition4dB}
    \centering
    \begin{ruledtabular}
      \begin{tabular}{lcccccc}
        & \multicolumn{2}{c}{\multirow{2}{*}{Input squeezing levels}}
        & \multirow{2}{*}{Transmissivity}
        & \red{Estimated output}
        & \multirow{2}{*}{Quality\tnote{*}}
        & \multirow{2}{*}{Photon detection rate\tnote{*}}
        \\
        & \multicolumn{2}{c}{} & & \red{squeezing level} & &
        \\
        & $e^{-2r_1}$ (dB) & $e^{-2r_2}$ (dB)
        & $T$
        & \red{$e^{-2r_{\text{est}}}$} (dB)
        & $W_{00} = \min W(x,p)$
        & (cps)\\
        \hline
        PS-4 & 4.14 & 0 & 0.97 & 3.40 & $-0.111 \pm 0.006$ & \num{9.96(4)e2} \\
        PS-5 & 4.24 & 0 & 0.95 & 3.35 & $-0.107 \pm 0.006$ & \num{1.895(6)e3} \\
        PS-6 & 4.35 & 0 & 0.93 & 3.24 & $-0.102 \pm 0.006$ & \num{2.496(8)e3} \\
        GPS-4 & 4.40 & $-0.39$ & 0.93 & 3.36 & $-0.110 \pm 0.006$ & \num{3.15(1)e3} \\
        GPS-5 & 4.70 & $-0.67$ & 0.89 & 3.30 & $-0.115 \pm 0.006$ & \num{5.51(2)e3} \\
        GPS-6 & 5.00 & $-0.94$ & 0.86 & 3.31 & $-0.112 \pm 0.005$ & \num{8.09(2)e3} \\
      \end{tabular}  
    \end{ruledtabular}
    \footnotesize
    \begin{tablenotes}
      \item[*] Shown with the standard errors.
    \end{tablenotes}
  \end{threeparttable}
  \end{minipage}
\end{table*}  

\section{Conclusion\label{sec:Conclusion}}
We numerically analyzed the generation of squeezed single-photon states by GPS with an on/off detector.
We derived a trade-off relationship between the qualities $W_{\text{out}}(0,0)$ and photon detection probabilities $P_{\text{on}}$ of the heralded states for both PS and GPS.
Our calculation shows that GPS can more efficiently generate states with comparable quality to PS, and it was experimentally proven for 2-dB- and 4-dB-squeezed single-photon states.
In fact, the generation rate improvement at \SI{2}{\deci\bel} exceeds one order of magnitude in our experiment.

Our theoretical and experimental studies reveal that we can boost the generation rates of the squeezed single-photon states using GPS, which can be realized only by a slight extension from the conventional PS.
\red{GPS is easily implementable because it only requires input squeezed states, which can be easily prepared deterministically, and an on/off detector in our case.}
This will accelerate the applications of squeezed single-photon states, which have been limited by the trade-off in PS, and also stimulate the progress of advanced CV QIP that requires non-Gaussian states.

\begin{acknowledgments}
  This work was partly supported by JST Grants No. JPMJFR223R, No. JPMJMS2064, and No. JPMJPF2221; JSPS KAKENHI Grants No. 22K20351, No. 23H01102, No. 23K13038, and No. 23K17300; the Canon Foundation; MEXT Leading Initiative for Excellent Young Researchers; and the Research Foundation for Opto-Science and Technology.
  H.T. and A.M. acknowledge financial support from The Forefront Physics and Mathematics Program to Drive Transformation (FoPM), WINGS Program, the University of Tokyo.
  The authors thank Takahiro Mitani for the careful proofreading of the manuscript.
\end{acknowledgments}

\appendix*
\section{\red{Relationship between negativity and fidelity}\label{sec:appendix}}
\red{
  We discuss the relationship between the negative value of the Wigner function at the origin $W_\text{out}(0,0)$ and the fidelity between the heralded state and the target squeezed single-photon state $F_{\text{out}}$.
  First, let us write the heralded states $\hat{\rho}_{\text{out}}$ in PS and GPS as
  \begin{align}
    \hat{\rho}_{\text{out}} = \sum_{n=1}^{\infty} P_n' \ket{\psi_n} \bra{\psi_n},
  \end{align}
  where $P_n '$ is the relative probability of $n$-photon detection when an ``on" signal appears and $\ket{\psi_n}$ is a quantum state heralded by $n$-photon detection.
  Here, $\ket{\psi_1}$ represents the target squeezed single-photon state because we assume the parameter condition where the target state is perfectly generated by single-photon detection.
  In this case, $W_\text{out}(0,0)$ and $F_{\text{out}}$ can be calculated as follows~\cite{agarwal2013quantum}.
  \begin{gather}
    \begin{aligned}
      W_\text{out}(0,0) &= \frac{1}{\pi} \sum_{n=1}^{\infty} \left(P_{2n} ' - P_{2n-1} '\right) \\
      &= \frac{1}{\pi} \left(1-2\sum_{n=1}^{\infty}P_{2n-1}' \right) \\
    \end{aligned}
    \\
    \begin{aligned}
      F_{\text{out}} &= \braket{\psi_1|\hat{\rho}_{\text{out}}|\psi_1} \\
      &= \sum_{n=1}^{\infty} P_{2n-1} '
      \left\Vert\braket{\psi_1|\psi_{2n-1}}\right\Vert ^2 
    \end{aligned}
  \end{gather}
  Here we used $\sum_{n=1}^{\infty} P_n ' = 1$ and $\braket{\psi_1|\psi_{2n}}=0\ (n=1,2,\cdots)$.
  The latter equation holds because the squeezed single-photon states $\ket{\psi_1}$ (superpositions of odd number states) and quantum states heralded by even number photon detection (superpositions of even number states) are orthogonal.
  When the detection probabilities of three or more photons are negligible, namely $P_{2n-1} '\approx0 \ (n=2,3,\cdots)$, the above equations are approximated as
  \begin{gather}
    W_\text{out}(0,0) \approx \frac{1}{\pi} \left(1-2P_1 '\right), \\
    F_\text{out} \approx P_1 ',
  \end{gather}
  and the following relationship is derived.
  \begin{align}
    F_\text{out} \approx \frac{1}{2} \left( 1- \pi  W_\text{out}(0,0)\right).
  \end{align}
  This relation indicates that $W_\text{out}(0,0)$ and $F_\text{out}$ are almost equivalent measures.
  In fact, most previous PS experiments~\cite{wenger_non-gaussian_2004,ourjoumtsev_generating_2006,neergaard-nielsen_generation_2006,wakui_photon_2007,dong_generation_2014,asavanant_generation_2017,takase_2022_generation} chose experimental condition where three or more photon-detection probabilities are negligibly small.
  Hence, if we suppose the practical experimental situations, minimizing $W_{\text{out}}(0,0)$ as in Sec.~\ref{sec:TheoryNumericalresults} almost corresponds to maximizing $F_{\text{out}}$.
  Even if we use the fidelity as an alternative quality indicator, our theoretical conclusion in this paper will remain unchanged.
}


\bibliography{PSvsGPS_arXivfinal}

\end{document}